\documentclass[10pt,letterpaper]{article}
\usepackage{spconf,amsmath,amssymb,booktabs,graphicx,float}
\usepackage[hidelinks]{hyperref}
\usepackage{orcidlink}
\hypersetup{pdftitle={Top-K Is Not a Budget for Hybrid Retrieval},pdfauthor={Chunran Zhang}}
\title{Top-K Is Not a Budget for Hybrid Retrieval}
\name{Chunran Zhang\,\orcidlink{0009-0005-8865-2090}}
\address{School of Computing and Artificial Intelligence\\
Southwest Jiaotong University, Chengdu, China\\
chronis@my.swjtu.edu.cn}

\begin{document}
\raggedbottom
\maketitle
\begin{abstract}
Modern hybrid retrieval for RAG typically fuses the Top-\(L\) results from dense and sparse retrievers, but a fixed truncation depth may not transfer across changing queries and corpora. Exact fusion removes the dependence on a fixed depth, yet completing a specified Top-\(K\) still incurs variable access costs. We present DiBud, which takes an access budget directly as input and incrementally certifies and returns an exact prefix of the RRF ranking over the full lists. Selective access increases certified output within the budget, while budgeted stopping bounds accesses per request. Experiments on five query sets reveal long-tailed costs for completing exact Top-20. At a budget of 2048 accesses, DiBud increases mean certified output within the first 100 positions by 7.86\% over balanced access. After budget calibration for 95\% quality retention, held-out queries retain 95.05\%--97.68\% of mean nDCG@20 while using 65.92\%--99.53\% fewer accesses than completing exact Top-20.

\end{abstract}
\section{Introduction}

\label{sec:intro}

Hybrid retrieval for retrieval-augmented generation (RAG) commonly combines the top \(L\) results from dense and sparse retrievers using reciprocal rank fusion (RRF)~\cite{elastic,azure,rrf}. A fixed window limits the number of entries accessed but also excludes contributions beyond that window. When a document appears in one ranked list but not in the other list's window, fusion assigns zero to the latter contribution. Yet absence from the window only means that the rank has not been observed; its contribution remains unknown. Replacing an unknown contribution with zero can change the output order even when the candidate set already contains every document in the top \(K\) under RRF over the full ranked lists. Thus, \(L\) is a retrieval-depth parameter that affects both retrieval quality and access cost.

In practice, \(L\) can be selected using historical queries to balance quality and cost. However, RAG systems serving open-ended requests over evolving knowledge bases continually encounter new queries and updated corpora. Both changes alter the ranked lists and the positions of cross-channel contributions, making it difficult for a previously selected \(L\) to preserve the original quality--cost tradeoff. Exact rank aggregation maintains bounds on unobserved contributions and accesses the ranked lists until the required results are determined~\cite{nra}. EAHR applies this approach to hybrid retrieval to obtain the exact top \(K\) in order without a predefined truncation depth~\cite{eahr}. Nevertheless, the number of accesses required can vary sharply across queries: resolving competition at just a few positions can require accessing entries far down the ranked lists.

Completing the same Top-\(K\) can require sharply different numbers of accesses across queries. Specifying \(K\) therefore does not provide a reliable bound on access cost. We instead take an access budget directly as input, limiting the total number of entries read from the two ranked lists. Within this budget, the system certifies results incrementally. When the budget is exhausted, it returns the certified prefix rather than continuing until a fixed number of results is obtained.

We introduce DiBud (Direct Budgeting) for retrieval under this direct access constraint. DiBud maintains bounds on unread contributions and incrementally appends certified positions~\cite{lara}. The output is an exact prefix: its documents and their order match the RRF ranking over the full lists. When the next position remains unresolved, selective access uses the current competition to choose which list to advance~\cite{snra}. Once the budget is exhausted, DiBud returns the accumulated prefix. Certification preserves the result semantics; the budget independently bounds the accesses required to produce that output. Selective reading aims to certify more results within the same budget without relaxing this guarantee.

We evaluate DiBud through replays on five query sets and five corpus snapshots. Fixed-depth experiments examine whether a selected \(L\) transfers across queries and corpus updates. Complete exact Top-20 runs measure how strongly access costs vary under the same output requirement. Equal-budget comparisons with balanced access assess certified output, while calibration on disjoint queries measures quality retention and access savings relative to completing exact Top-20. Together, these experiments test whether directly bounding accesses can retain useful exact prefixes while avoiding the long-tail cost of a fixed result count.

\section{DiBud: Budgeted Exact-Prefix Fusion}

\label{sec:method}

\begin{figure}[H]
\centering
\includegraphics[width=\columnwidth]{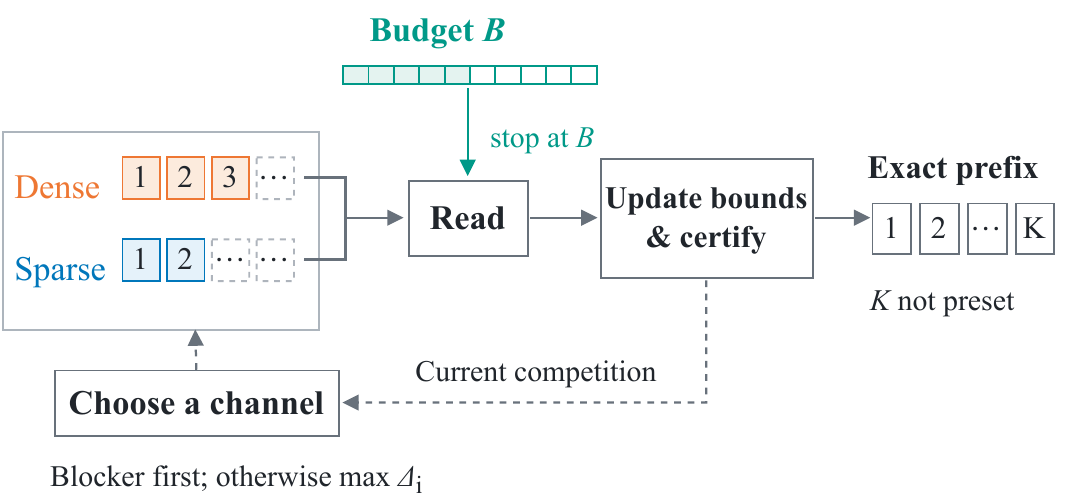}
\caption{DiBud overview. The budget directly limits reads, while current competition guides channel selection. Certification determines the returned prefix.}
\label{fig:method}
\end{figure}

Given a query, a corpus snapshot, and a read budget \(B\), DiBud reads the dense and sparse rankings on demand and returns a certified prefix of their complete RRF ranking. The budget determines how much can be read; certification determines how much can be returned (Fig.~\ref{fig:method}).

\subsection{Budget and output}

\label{budget-and-output}

Let \(r_i(x)\) denote the one-based rank of document \(x\) in channel \(i\in\{D,S\}\), and let \(g(r)=1/(c+r)\) for rank constant \(c\geq 0\). Set \(r_i(x)=\infty\) and \(g(\infty)=0\) if \(x\) is outside that channel's support. The complete fusion score is \begin{equation}
F(x)=\sum_{i\in\{D,S\}}g(r_i(x)).
\label{eq:score}
\end{equation} Sorting the union of the channel supports by decreasing \(F(x)\), with a fixed document-identifier order to break ties, defines the complete sequence \(T^\star\).

Both channels supply resumable exact ranking prefixes from the same snapshot. If \(d_D\) and \(d_S\) are their read depths and \(A\) is the output sequence, the required contract is \begin{equation}
d_D+d_S\leq B,\qquad A=T^\star_{1:K},\quad K=|A|.
\label{eq:contract}
\end{equation} Here \(B\) is the input limit; \(K=|A|\) is the certified output size, not a preset count. Exactness applies to these returned positions. If none is certified, \(A\) is empty. Reading the same document from both channels counts as two accesses.

\subsection{Prefix certification}

\label{prefix-certification}

Let \(S_i\) be the documents already read from channel \(i\). An unread contribution in that channel is bounded by \begin{equation}
u_i=\begin{cases}
 0,&\text{if confirmed exhausted},\\
 g(d_i+1),&\text{otherwise}.
 \end{cases}
\label{eq:unread}
\end{equation} For an observed document \(x\), the score bounds are \begin{align}
\ell(x)&=\sum_{i:x\in S_i}g(r_i(x)),\label{eq:lower}\\
h(x)&=\ell(x)+\sum_{i:x\notin S_i}u_i.\label{eq:upper}
\end{align} A document unseen in both channels has score at most \(h_{\varnothing}=u_D+u_S\).

Among observed documents not yet emitted, select \(x\) with the highest lower bound, breaking ties by the fixed identifier order. Certify \(x\) as the next output when its lower bound exceeds every remaining competitor's upper bound and \(h_{\varnothing}\). Equality with an observed competitor is allowed only when \(x\) wins the identifier tie; comparison with the unseen bound remains strict. Append \(x\) and repeat until the next position cannot be certified.

The bounds ensure \(\ell(x)\leq F(x)\leq h(x)\). Each emitted document therefore precedes every remaining document, including any not yet observed. Applying this argument at each output position establishes the prefix equality in~\eqref{eq:contract}. Each position can be certified without resolving later positions, so the prefix grows incrementally and remains valid when the budget ends.

\subsection{Reading and stopping}

\label{reading-and-stopping}

When the next position cannot be certified, let \(y\) be the observed competitor with the highest upper bound, excluding \(x\) and the emitted prefix and breaking ties by identifier. If \(h(y)\geq h_{\varnothing}\) and \(y\) is missing a contribution from exactly one channel, advance that channel. Reading may reveal \(y\) and determine its missing contribution. Otherwise, advancing the read depth lowers its contribution bound. Both outcomes help determine whether \(y\) still blocks the next position.

Otherwise, advance the channel whose next batch offers the larger reduction in the unread bound. For a batch of \(b\) entries in a channel that remains open after the read, this reduction is \begin{equation}
\Delta_i=g(d_i+1)-g(d_i+b+1).
\label{eq:drop}
\end{equation} If exhaustion is established, the new bound is zero. Exhausted channels are skipped, and equal reductions are resolved by a fixed channel order (dense first). This rule also initializes reading when no document has been observed.

After each batch, update the bounds and extend the certified prefix. The batch size is capped by the remaining budget. Complete these updates after the final allowed read, then return the accumulated output when the budget is consumed or both channels are exhausted. The schedule affects how many positions can be certified within the budget; the certification rule preserves the exactness of every returned position.

\begin{figure*}[!t]
\centering\includegraphics[width=\textwidth]{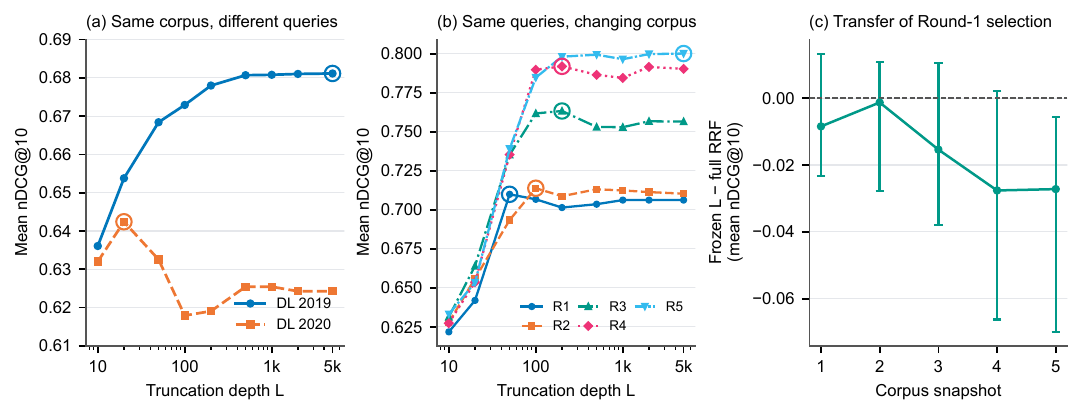}
\caption{Depth sensitivity and transfer. (a) Different queries on the same corpus. (b) The same queries across snapshots, using each round's judgments. Circles mark the best tested depths. (c) Held-out transfer of Round-1 depths. Error bars show 95\% nested bootstrap intervals with depth reselection and query identities preserved across rounds.}
\label{fig:transfer}\end{figure*}

\section{Experiments}

\label{sec:experiments}

\subsection{Setup}

\label{setup}

We evaluate 770 queries from five query sets---TREC-DL 2019/2020, NFCorpus, SciFact, and TREC-COVID---and the same 30 queries across five TREC-COVID snapshots~\cite{dl19,dl20,nfcorpus,scifact,covid}. We replay exact rankings from bge-small-en-v1.5~\cite{bge} and BM25~\cite{bm25} (\(k_1=0.9\), \(b=0.4\)), using equal-weight RRF with contributions \(1/(59+r)\) and deterministic tie breaking.

DiBud and Balanced use the same certification rules and access budgets, ranging from 128 to 10000. Both read one entry at a time; Balanced alternates between the two lists. We count certified results up to 20 and 100, and use completed exact Top-20 as the cost and quality reference.

Static replays use reconstructed full rankings: all 770 queries complete exact Top-20. Temporal replays use frozen snapshot rankings without treating saved-list boundaries as exhaustion. Results are averaged within each query set, then equally across sets. Reported 95\% confidence intervals use 2000 query bootstrap samples.

\subsection{Transfer of fixed depths}

\label{transfer-of-fixed-depths}

We evaluate \(L\in\{10,\allowbreak20,\allowbreak50,\allowbreak100,\allowbreak200,\allowbreak500,\allowbreak1000,\allowbreak2000,\allowbreak5000\}\). On the same MS MARCO corpus, increasing \(L\) from 20 to 5000 raises nDCG@10 from 0.6538 to 0.6812 for TREC-DL 2019, but lowers it from 0.6424 to 0.6243 for TREC-DL 2020 (Fig.~\ref{fig:transfer}a). Changing queries reverses the benefit of a deeper window.

For the same 30 queries across five TREC-COVID snapshots, the best tested depths are 50, 100, 200, 200, and 5000 (Fig.~\ref{fig:transfer}b). We then test whether an earlier selection transfers: five-fold calibration selects \(L\) using Round-1 queries and freezes it for held-out queries across all rounds. The nDCG@10 difference from full RRF changes from \(-0.0085\) in Round 1 to \(-0.0272\) in Round 5 (95\% CI: \([-0.0699,-0.0058]\); Fig.~\ref{fig:transfer}c). A depth selected on earlier data does not necessarily maintain its relative effectiveness as the corpus changes.

\subsection{Fixed-K access cost}

\label{fixed-k-access-cost}

Completing exact Top-20 is inexpensive for most queries but costly in the tail (Table~\ref{tab:cost}). On TREC-DL 2019, half the queries finish within 311 accesses, whereas the P95 reaches 196410 and the maximum reaches 3533518---approximately 632 and 11362 times the median. The same pattern appears on TREC-DL 2020 and TREC-COVID, where the P95 reaches 80 and 22 times the median, respectively. A \(K\) that is affordable for most queries can therefore be expensive for others within the same query set. This variability motivates specifying the access budget directly and determining the output size during execution, without a preset \(K\).

\begin{table}[H]
\centering
\caption{Accesses to complete exact Top-20 under unit-access DiBud. All 770 queries complete.}
\label{tab:cost}
\small\setlength{\tabcolsep}{3pt}
\begin{tabular}{lrrrr}
\toprule
Query set & $n$ & Median & P95 & Maximum \\
\midrule
TREC-DL 2019 & 43 & 311 & 196410 & 3533518 \\
TREC-DL 2020 & 54 & 444 & 35562 & 211169 \\
NFCorpus & 323 & 308 & 2685 & 5035 \\
SciFact & 300 & 325 & 3362 & 5011 \\
TREC-COVID & 50 & 2544 & 56363 & 91290 \\
\bottomrule
\end{tabular}
\end{table}

\subsection{Certified output under equal budgets}

\label{certified-output-under-equal-budgets}

We first isolate the effect of list selection by holding the budget and certification rule fixed. With the same budget of 2048 accesses, DiBud certifies more results than Balanced on all five query sets (Table~\ref{tab:yield}). Averaged equally across sets, output within the first 100 positions increases from 41.77 to 45.05, a 7.86\% gain; within the first 20, it increases from 18.07 to 18.18. Selective reading therefore increases the number of exact positions obtained from the same access allowance.

\begin{table}[H]
\centering
\caption{Mean certified output at $B=2048$. Both methods use unit accesses and the same certifier; only list selection differs.}
\label{tab:yield}
\small
\setlength{\tabcolsep}{3pt}
\begin{tabular}{lrrrr}
\toprule
& \multicolumn{2}{c}{Cap 100} & \multicolumn{2}{c}{Cap 20} \\ Query set & Balanced & DiBud & Balanced & DiBud \\
\midrule
TREC-DL 2019 & 41.86 & 46.19 & 17.77 & 17.91 \\
TREC-DL 2020 & 34.91 & 38.31 & 18.33 & 18.35 \\
NFCorpus & 62.06 & 63.71 & 19.24 & 19.28 \\
SciFact & 43.55 & 47.26 & 19.44 & 19.57 \\
TREC-COVID & 26.48 & 29.80 & 15.58 & 15.78 \\
\bottomrule
\end{tabular}
\end{table}

\subsection{Quality and access cost relative to exact Top-20}

\label{quality-and-access-cost-relative-to-exact-top-20}

We next examine the cost of requiring a fixed output count. We compare two stopping conditions on the same access trajectory: complete exact Top-20, or stop earlier when the budget is exhausted. Five-fold calibration selects the smallest budget retaining 95\% of the reference mean nDCG@20 and applies it to held-out queries. Missing output positions contribute zero gain.

\begin{table}[H]
\centering
\caption{Held-out results after 95\% budget calibration. Retention is the ratio of mean nDCG@20; savings compare total accesses with completed exact Top-20.}
\label{tab:quality}
\small\setlength{\tabcolsep}{3pt}
\begin{tabular}{lrr}
\toprule
Query set & Retained (\%) & Fewer accesses (\%) \\
\midrule
TREC-DL 2019 & 95.05 & 99.53 \\
TREC-DL 2020 & 95.89 & 93.50 \\
NFCorpus & 95.81 & 65.92 \\
SciFact & 97.68 & 83.41 \\
TREC-COVID & 95.85 & 69.71 \\
\bottomrule
\end{tabular}
\end{table}

Across the five query sets, budgeted stopping retains 95.05\%--97.68\% of mean nDCG@20 while reducing accesses by 65.92\%--99.53\% (Table~\ref{tab:quality}). Both stopping conditions follow the same access trajectory, so these savings come from stopping earlier. The 99.53\% reduction on TREC-DL 2019 reflects the high cost of completing tail queries. These results show that completing all 20 positions is not necessary to retain most of the measured retrieval quality. Directly limiting accesses and allowing the output size to vary preserves most of that quality while avoiding much of the completion cost.

\section{Conclusion}

\label{conclusion}

Changing queries and corpora make it difficult for a fixed Top-\(L\) depth to preserve its quality--cost tradeoff in RAG hybrid retrieval. Exact fusion removes this fixed cutoff, but a prescribed Top-\(K\) still leaves access cost dependent on the query. DiBud takes an access budget directly as input and incrementally returns certified results, determining the output size during execution.

Experiments on five query sets show that selective reading certifies more results than balanced reading under equal budgets. After calibration for 95\% quality retention, budgeted stopping retains 95.05\%--97.68\% of mean nDCG@20 on held-out queries while reducing accesses by 65.92\%--99.53\%. The long-tailed costs of completing exact Top-20 explain why direct control is needed: fixing the result count leaves the required accesses free to vary across queries. Top-\(K\) is therefore not a budget for hybrid retrieval.

\section{Acknowledgments}

\label{acknowledgments}

OpenAI Codex assisted with drafting and editing the abstract and Sections 1--4 from author-provided material, as well as implementing the experimental code.

\bibliographystyle{IEEEbib}
\bibliography{references}

\begin{thebibliography}{10}

\bibitem{elastic}
{Elastic},
\newblock ``Reciprocal rank fusion,'' \url{https://www.elastic.co/docs/reference/elasticsearch/rest-apis/reciprocal-rank-fusion},
\newblock Accessed September 10, 2026.

\bibitem{azure}
{Microsoft},
\newblock ``Hybrid search scoring ({RRF})---{Azure AI Search},'' \url{https://learn.microsoft.com/en-us/azure/search/hybrid-search-ranking},
\newblock Accessed September 11, 2026.

\bibitem{rrf}
Gordon~V. Cormack, Charles L.~A. Clarke, and Stefan Buettcher,
\newblock ``Reciprocal rank fusion outperforms condorcet and individual rank learning methods,''
\newblock in {\em Proceedings of the 32nd International ACM SIGIR Conference on Research and Development in Information Retrieval}, 2009, pp. 758--759.

\bibitem{nra}
Ronald Fagin, Amnon Lotem, and Moni Naor,
\newblock ``Optimal aggregation algorithms for middleware,''
\newblock {\em Journal of Computer and System Sciences}, vol. 66, no. 4, pp. 614--656, 2003.

\bibitem{eahr}
Chunran Zhang,
\newblock ``Exact adaptive hybrid retrieval without fixed top-{L} cutoffs,''
\newblock {\em arXiv preprint arXiv:2608.07152}, 2026.

\bibitem{lara}
Nikos Mamoulis, Kit~Hung Cheng, Man~Lung Yiu, and David~W. Cheung,
\newblock ``Efficient aggregation of ranked inputs,''
\newblock in {\em Proc. IEEE International Conference on Data Engineering (ICDE)}, 2006, p.~72.

\bibitem{snra}
Jing Yuan, Guang-Zhong Sun, Ye~Tian, Guoliang Chen, and Zhi Liu,
\newblock ``Selective-{NRA} algorithms for top-$k$ queries,''
\newblock in {\em Advances in Data and Web Management (APWeb/WAIM)}, 2009, pp. 15--26.

\bibitem{dl19}
Nick Craswell, Bhaskar Mitra, Emine Yilmaz, Daniel Campos, and Ellen~M. Voorhees,
\newblock ``Overview of the {TREC} 2019 deep learning track,''
\newblock {\em arXiv preprint arXiv:2003.07820}, 2020.

\bibitem{dl20}
Nick Craswell, Bhaskar Mitra, Emine Yilmaz, and Daniel Campos,
\newblock ``Overview of the {TREC} 2020 deep learning track,''
\newblock {\em arXiv preprint arXiv:2102.07662}, 2021.

\bibitem{nfcorpus}
Vera Boteva, Demian Gholipour, Artem Sokolov, and Stefan Riezler,
\newblock ``A full-text learning to rank dataset for medical information retrieval,''
\newblock in {\em Advances in Information Retrieval}. 2016, pp. 716--722, Springer.

\bibitem{scifact}
David Wadden, Shanchuan Lin, Kyle Lo, Lucy~Lu Wang, Madeleine van Zuylen, Arman Cohan, and Hannaneh Hajishirzi,
\newblock ``Fact or fiction: Verifying scientific claims,''
\newblock in {\em Proceedings of the 2020 Conference on Empirical Methods in Natural Language Processing}. 2020, pp. 7534--7550, Association for Computational Linguistics.

\bibitem{covid}
Kirk Roberts, Tasmeer Alam, Steven Bedrick, Dina Demner-Fushman, Kyle Lo, Ian Soboroff, Ellen Voorhees, Lucy~Lu Wang, and William~R. Hersh,
\newblock ``Searching for scientific evidence in a pandemic: An overview of {TREC-COVID},''
\newblock {\em Journal of Biomedical Informatics}, vol. 121, pp. 103865, 2021.

\bibitem{bge}
{BAAI},
\newblock ``{bge-small-en-v1.5}: Model card,'' \url{https://huggingface.co/BAAI/bge-small-en-v1.5},
\newblock Accessed September 10, 2026.

\bibitem{bm25}
Stephen Robertson and Hugo Zaragoza,
\newblock ``The probabilistic relevance framework: {BM25} and beyond,''
\newblock {\em Foundations and Trends in Information Retrieval}, vol. 3, no. 4, pp. 333--389, 2009.

\end{thebibliography}
\end{document}